\documentclass[openacc]{rsproca_new}

\usepackage{siunitx}
\usepackage{graphicx}
\usepackage{caption}
\usepackage{subcaption}
\usepackage[dvipsnames]{xcolor}
\usepackage{wrapfig}
\usepackage{comment}
\usepackage{float}
\usepackage{amsmath}
\usepackage{adjustbox}

\jname{}
\Journal{}

\begin{document}

\title{Entanglement by design: Symmetry-guided periodic helical assemblies}

\author{
Myfanwy E. Evans$^{1}$}

\address{$^{1}$
Universität Potsdam, Institut für Mathematik\\
14476 Potsdam, Germany\\
Email Address: evans@uni-potsdam.de}

\subject{Biomaterials, Biomathematics}

\keywords{Weave, tangle, knot, periodic nets, symmetry}

\corres{Myfanwy E. Evans\\
\email{evans@uni-potsdam.de}}

\begin{abstract}

In this paper, a selection of elegant, highly symmetric examples of three-periodic tangled nets and filaments are presented. They are constructed \textit{via} familiar crystal nets using edges as geometric scaffolds for $n$-fold helical windings. Rather than providing a complete classification, this gallery of examples highlights recurring geometric motifs, offering insight into how periodic tangles are organised in crystalline, molecular, and biological systems.
\end{abstract}


\begin{fmtext}

\section{Introduction}

Triply-periodic tangled geometry underlies the structure of a wide range of natural and synthetic systems, from crystalline frameworks \cite{mof_2001} and molecular assemblies \cite{zhang2022} to filament packings and biological tissues \cite{skin_1,Jessop2025WovenGyroids}. The study of such diverse materials requires a unifying language, where geometry has emerged as an important foundation. Early geometric approaches to crystal chemistry established that crystalline structures can be understood as idealised networks embedded in three-dimensional space \cite{Wells:a01232}, and that many structural features are governed as much by symmetry and spatial embedding as by chemical composition. Within this geometric viewpoint, it was later shown that complex three-dimensional periodic entanglements can be generated systematically by mapping hyperbolic line patterns into Euclidean space \cite{HydeOguey2000}, revealing entanglement as a natural consequence of geometry and symmetry. This perspective has been formalised through graph-theoretic descriptions of crystal nets as embedded periodic graphs, providing a rigorous framework for comparing and classifying crystalline networks independently of chemical realisation \cite{delgado-friedrichs2005crystal}.

A central theme emerging in this context is that entanglement is not an incidental complication of periodic structure, but a natural and generative feature of three-dimensional geometry
\cite{HydeOguey2000,myf2011_entanglement_graphs,periodic_ent_I,Evans:eo5020}. Periodic nets and filaments may interweave, thread, or knot while preserving long-range order, giving rise to structures that are simultaneously crystalline and topologically non-trivial. Such configurations extend classical knot and link theory into the periodic setting, revealing deep connections between symmetry, topology, and spatial embedding \cite{alexandrov2011,nets_MOFs,Power:ib5087,sym14040822,tesselate_decussate,okeeffe_treacy_2025_3periodic_weavings}. These differences in tangling are not merely aesthetic: they influence pore structure, mechanical response, and transport properties in crystalline materials, such as coordination networks \cite{CARLUCCI2003247,mof_2001,liu2016,zhang2022}, and they determine how filaments pack, slide, or resist deformation in biological systems \cite{skin_1,PhysRevLett.112.038102}. 

\end{fmtext}


\maketitle

\newpage

One particular route to symmetric tangling involves symmetric $n$-fold helices as fundamental tangling motifs, which are then joined together into more extended tangled patterns. In this setting, spatial graphs describe the arrangement of these helices in space and their mutual connections; graph edges represent central helical axes, and graph vertices the connectivity between adjacent helices. Such an approach has been applied to platonic polyhedral scaffolds to generate highly symmetric tangled platonic polyhedra \cite{HydeEvans2022TangledPlatonic}, and to entangled, 2-periodic honeycomb networks \cite{EvansHyde2022SymmetricHoneycomb}. Taken together, these papers build the foundations of a unifying theory in which tangling becomes a symmetry-preserving design principle applicable across finite and periodic networks. Closely related ideas of wound structures on polyhedra have also been explored theoretically, inspired by molecular and DNA nanostructures \cite{Hu2009PlatonicPolyhedralLinks, Guo2019ChiralityTriangularPrismLinks}, as well as in molecular self-assembly \cite{Sawada2019MetalPeptideTopologicalFrameworks, Domoto2020CoordinationPolyhedraEntangledFaces}.

\begin{figure}[h]
\centering
\includegraphics[width=.35\linewidth]{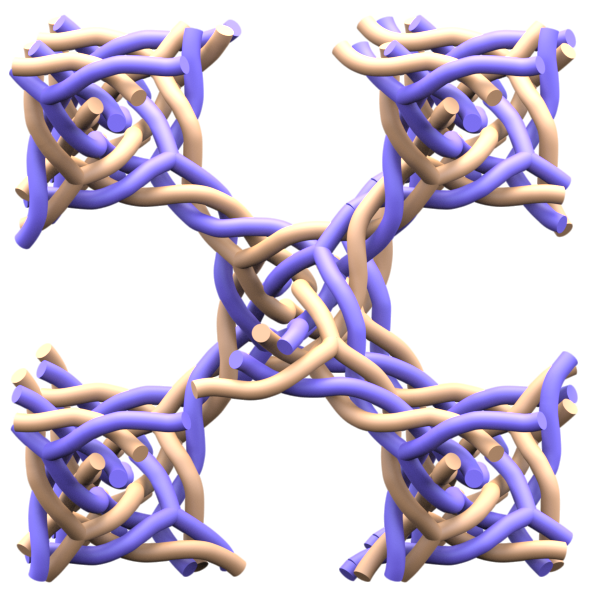}  
\caption{An example of an elegant entanglement of two like-handed \textbf{srs} nets. The structure has very high symmetry and a complicated tangling pattern.}
\label{fig:intro}
\end{figure}

In this paper, I build directly on this construction principle to present a curated selection of highly symmetric tangles and weavings that wind around three-periodic nets. Treating the edges of a parent net as scaffolds for $n$-fold helices, we illustrate how periodic entanglements arise under strong constraints imposed by symmetry and connectivity. The examples include both tangled graphs, in which strands meet at vertices, and filament weavings, in which 1-dimensional strands remain disjoint yet tangled. An example of such an elegant, highly symmetric, tangled net structure is shown in Figure \ref{fig:intro}, in this case two like-handed interwoven \textbf{srs} nets \cite{rcsr}, which we will discuss later in the paper. 

\begin{figure}[h]
\centering
\begin{subfigure}{.24\textwidth}
    \centering
    \includegraphics[width=.95\linewidth]{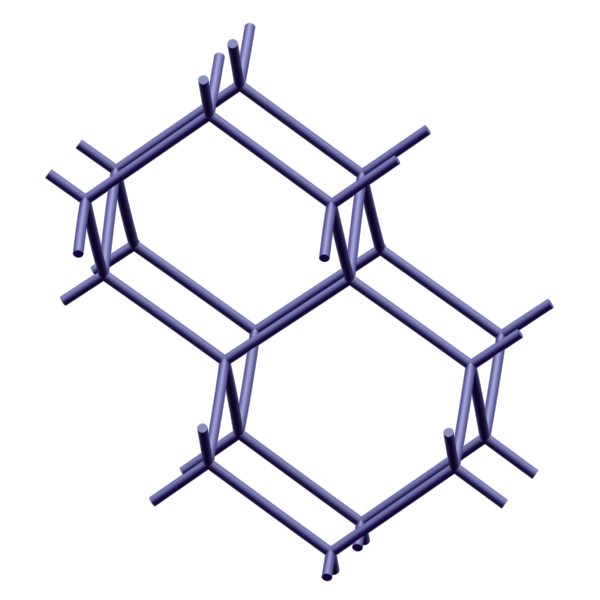}  
    \caption{\textbf{dia} net}
    \label{SUBFIGURE LABEL 1}
\end{subfigure}
\begin{subfigure}{.24\textwidth}
    \centering
    \includegraphics[width=.95\linewidth]{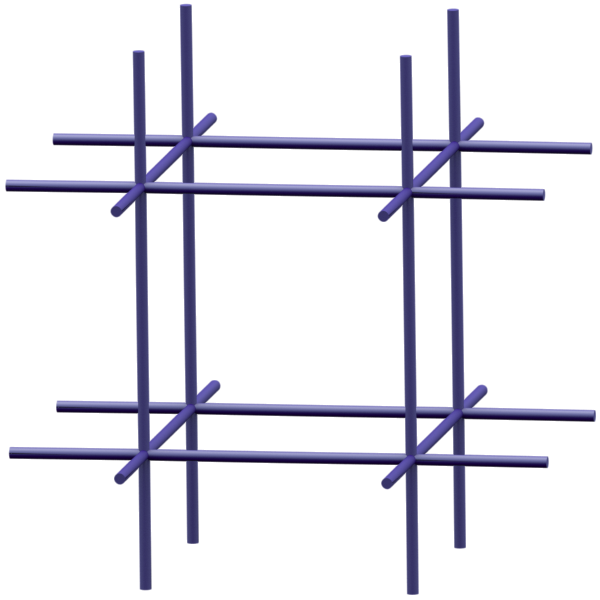}  
    \caption{\textbf{pcu} net}
    \label{SUBFIGURE LABEL 2}
\end{subfigure}
\begin{subfigure}{.24\textwidth}
    \centering
    \includegraphics[width=.95\linewidth]{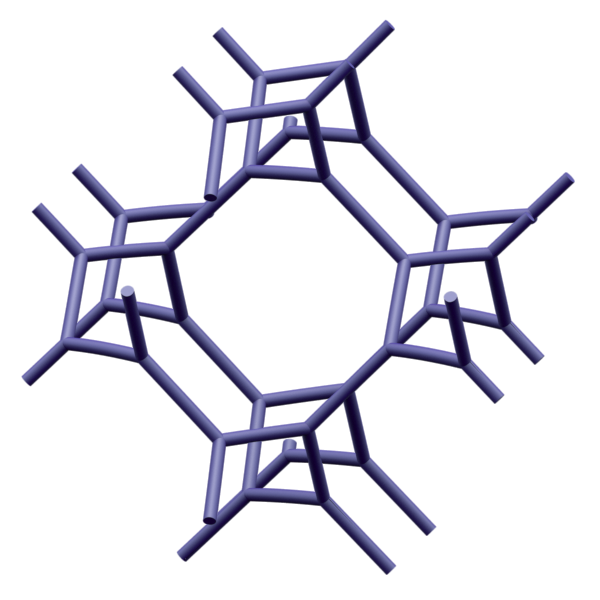}  
    \caption{\textbf{srs} net}
    \label{SUBFIGURE LABEL 3}
\end{subfigure}
\caption{The three candidate scaffold nets used in the paper for the contruction of periodic tangles, with their names as taken from the RCSR database \cite{rcsr} shown below each structure.}
\label{fig:nets}
\end{figure}

Focusing on the least genus, highest symmetry regular nets \textbf{srs}, \textbf{dia}, and \textbf{pcu} \cite{delgado-friedrichs2003_regular_quasiregular} (Figure \ref{fig:nets}, nomenclature from the RCSR database \cite{rcsr}) as scaffolds, we show that the geometric requirements imposed by the $n$-fold symmetry of the $n$-fold helices alongside periodicity lead to an elegant set of tangled structures. Of the three underlying scaffold nets, the \textbf{srs} net has 2-fold rotational symmetry along the axes of its edges, the \textbf{dia} net has 3-fold symmetry along its edges, and the \textbf{pcu} has 4-fold symmetry. This allows helices with $2n$ strands to preserve all (rotational) symmetries of the \textbf{srs} net when place along its edges, $3n$ strands for the \textbf{dia} net, and $4n$ strands for the \textbf{pcu} net. These constraints lead to a striking economy of form: a small number of underlying nets generate a rich but highly structured family of periodic tangles, many of which correspond to known interpenetrated frameworks, molecular weavings, or invariant rod packings \cite{cubic_rod_packings}.

The spatial organisation inherited from the underlying net and the helical geometry allows the description of each resulting tangled structure using a short symbolic code that records the helical pitch and the network connectivity \cite{HydeEvans2022TangledPlatonic,EvansHyde2022SymmetricHoneycomb}. This notation gives a direct and reproducible geometric description of the structure in three-dimensional space.

By presenting these structures as a gallery of illustrative examples rather than a complete classification, this work aims to clarify how three-dimensional symmetry, geometry and topology conspire to produce ordered complexity. Beyond their mathematical interest, such tangled structures offer insight into the geometric principles governing crystalline frameworks, molecular assemblies, and biological materials, all settings where geometry plays a central role.

\section{Assembling double helices into tangled structures}

We begin by discussing assemblies of double helices into extended periodic tangled structures. Of the three candidate scaffold nets, only the \textbf{srs} satisfies the symmetry restrictions of the 2-fold rotational symmetry along the axis of the double helix, as discussed above. The \textbf{srs} net is a periodic graph of genus-3 with four distinct vertices and six distinct edges in a periodic unit cell. Along each of these six edges, we can place double helices with a particular pitch, described by $\frac{t}{2}$, where the 2 strands (denominator) make a twist of $t\pi$ in the helix. The index $\frac{2}{2}$ indicates a complete turn $2\pi$ of the double helix. Prioritising symmetry dictates that all six double helices arranged on the \textbf{srs} net should have the same pitch, which gives a \textit{tangle index} of $\{\frac{t}{2}\}^6$.

\begin{figure}[h]
\centering
\begin{subfigure}{.27\textwidth}
    \centering
    \includegraphics[width=.95\linewidth]{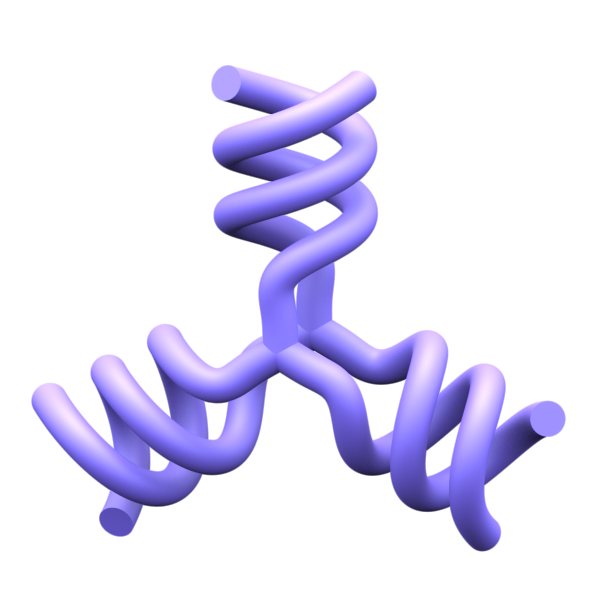}  
    \caption{}
    \label{SUBFIGURE LABEL 3}
\end{subfigure}
\begin{subfigure}{.27\textwidth}
    \centering
    \includegraphics[width=.95\linewidth]{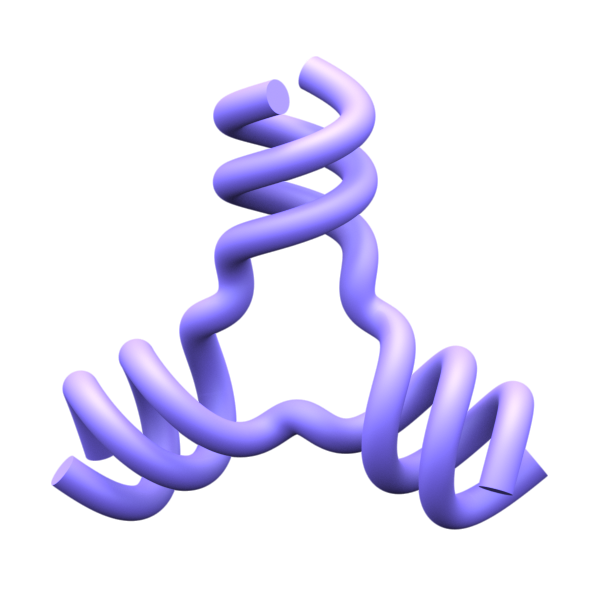}  
    \caption{}
    \label{SUBFIGURE LABEL 4}
\end{subfigure} \hfill
\caption{Two possible maximal symmetry closures of three double helices arranged around a symmetric degree-3 vertex. The open ends of the helices can either join together at two vertices located at the 3-fold symmetry axes (a), or pairwise through 2-fold symmetry axes (b).}
\label{fig:3-closures}
\end{figure}

With double helices arranged along the edges of the net, the open ends of these helices need to be connected together to create a closed structure. In the case of double helices, this can be done in two distinct ways that preserve the maximal symmetry of the underlying \textbf{srs} net. These two closures, shown in Figure \ref{fig:3-closures}, result in either network-like or infinite filament components, which trace through sequences of double helices in space. We describe these as the \textit{net closure} or the \textit{weave closure} of the helices.

The final consideration in the construction is the constraints on possible values of $t$ in the tangle indices. Neighbouring 3-fold vertices of the \textbf{srs} net are not coplanar, thus some torsion ($0.4\pi$) is already present in the edge of the net; to ensure that helices can join seamlessly at both ends of the helix as required, the twist of the helices must be $t+0.6$ (where $t$ is an integer) on the left-handed \textbf{srs} net (equivalent to a twist of $-0.4\pi$, shifted by an integer), or $t-0.6$ on the right-handed \textbf{srs} net. This $0.4\pi$ twist along each edge is slightly larger than the expected ca. $70.5^\circ$ (the local angle twist of an edge in \textbf{srs}), and comes from the the total twist around the 10-edge cycles of the net ($720^\circ$) distributed across each of the 10 edges ($72^\circ = 0.4\pi$). This gives an exact expression for the inherent twist in the edges required to connect threads seamlessly. In this paper, we present only left-handed \textbf{srs} structures for simplicity, but their mirror images can be constructed analogously. 

\begin{figure}[h]
\centering
\begin{subfigure}{.31\textwidth}
    \centering
    \includegraphics[width=.95\linewidth]{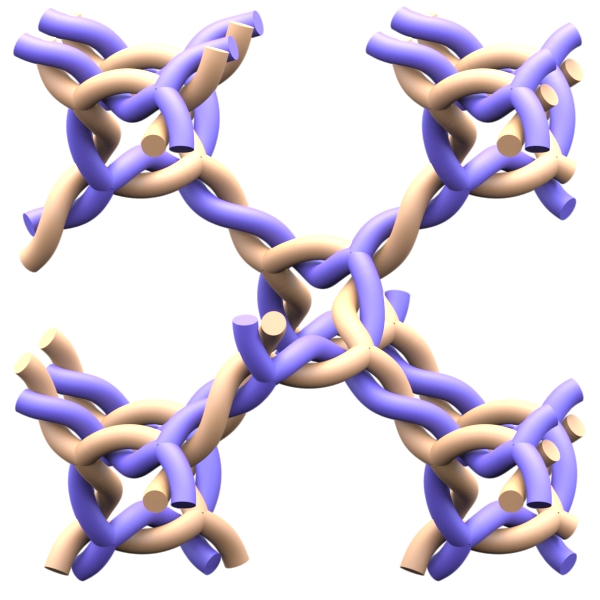}  
    \caption{\textbf{srs} net $\{\frac{\overline{1.4}}{2} \}^6$}
    \label{SUBFIGURE LABEL 1}
\end{subfigure}
\begin{subfigure}{.31\textwidth}
    \centering
    \includegraphics[width=.95\linewidth]{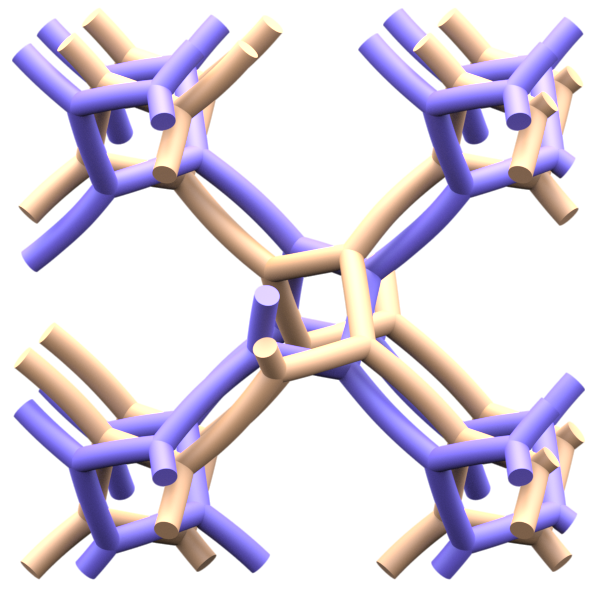}  
    \caption{\textbf{srs} net $\{\frac{\overline{0.4}}{2} \}^6$}
    \label{SUBFIGURE LABEL 2}
\end{subfigure}\\
\begin{subfigure}{.31\textwidth}
    \centering
    \includegraphics[width=.95\linewidth]{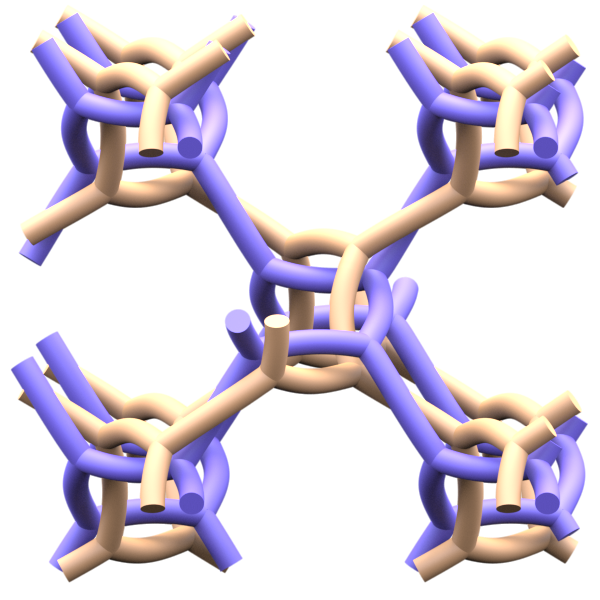}  
    \caption{\textbf{srs} net $\{\frac{0.6}{2} \}^6$}
    \label{SUBFIGURE LABEL 3}
\end{subfigure}
\begin{subfigure}{.31\textwidth}
    \centering
    \includegraphics[width=.95\linewidth]{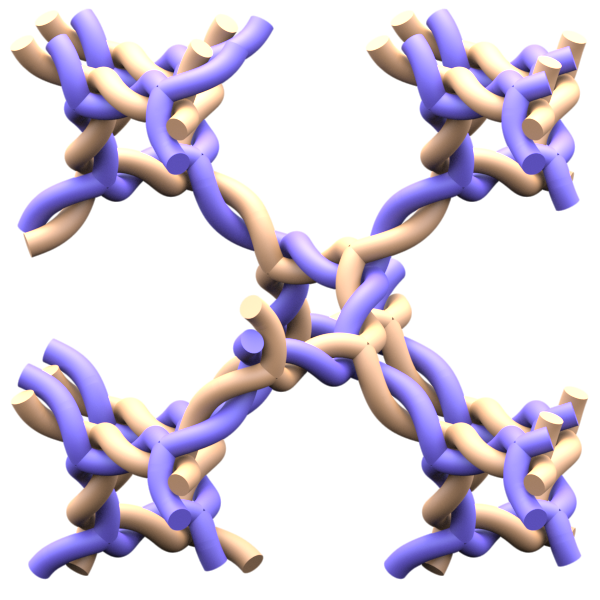}  
    \caption{\textbf{srs} net $\{\frac{1.6}{2} \}^6$}
    \label{SUBFIGURE LABEL 4}
\end{subfigure} \hfill
\caption{Four distinct double-helix tangles hung from the \textbf{srs} net with net closures. In all four cases, the resulting 3-dimensional structure is a tangling of two \textbf{srs} nets (both left-handed, shown in purple and yellow), with each of the different helix pitch choices leading to different tangling of the nets in space. Their tangle indices are shown below each structure. All of the structures have $I4_{1}32$ symmetry.}
\label{fig:2helices_net}
\end{figure}

Figure \ref{fig:2helices_net} shows a set of simplest examples arising on the left-handed \textbf{srs} net with a net closure of the double helices. Examples are shown for helical twists of $-1.4\pi$ ($t=-2$), $-0.4\pi$ ($t=-1$), $0.6\pi$ ($t=0$), and $1.6\pi$ ($t=1$). In all cases, the resulting structures contain two distinct but like-handed curvilinear \textbf{srs} nets in space, with each of the twists giving a different mutual tangling of the nets. The design of these structures on the left-handed srs net leads to both \textbf{srs} components of the tangled structure also being left-handed. Three of these examples have been described and observed previously; \textbf{srs} net $\{\frac{\overline{0.4}}{2} \}^6$ and \textbf{srs} net $\{\frac{0.6}{2} \}^6$ were previously described in \cite{HydeOguey2000}, \cite{CastleEvansHydeRamsdenRobins2012} and \cite{periodic_ent_I} from the perspective of hyperbolic tilings. This new construction allows us to extend to a wider variety of tangled structures in a systematic way. The structure \textbf{srs} net $\{\frac{\overline{0.4}}{2} \}^6$ has also be described in various contexts describing physical structure, including in a microporous metal-organic framework \cite{CussenClaridgeRosseinskyKepert2002} and more recently in a liquid crystal polymer \cite{TangEtAl2025Angew}. In addition, the structure \textbf{srs} net $\{\frac{\overline{1.4}}{2} \}^6$ is related to a molecular structure in a metal-peptide assembly \cite{Inomata2025M60L60Capsid}.

\begin{figure}[h]
\centering
\begin{subfigure}{.32\textwidth}
    \centering
    \includegraphics[width=.95\linewidth]{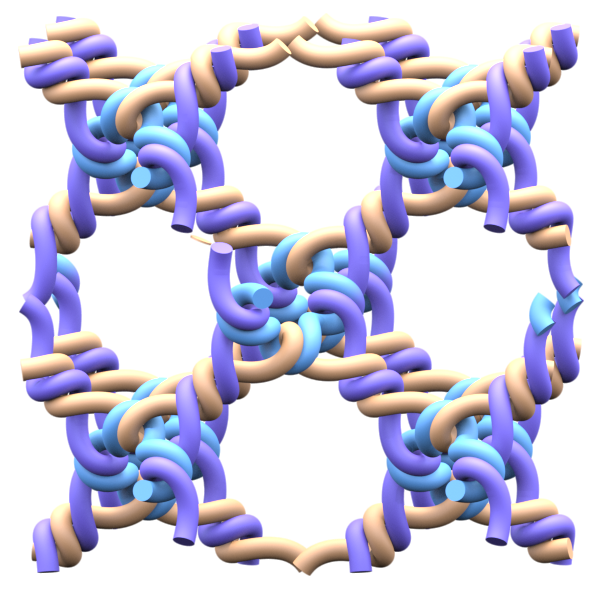}  
    \caption{\textbf{srs} weave $\{\frac{\overline{2.4}}{2} \}^6$}
    \label{SUBFIGURE LABEL 1}
\end{subfigure}
\begin{subfigure}{.32\textwidth}
    \centering
    \includegraphics[width=.95\linewidth]{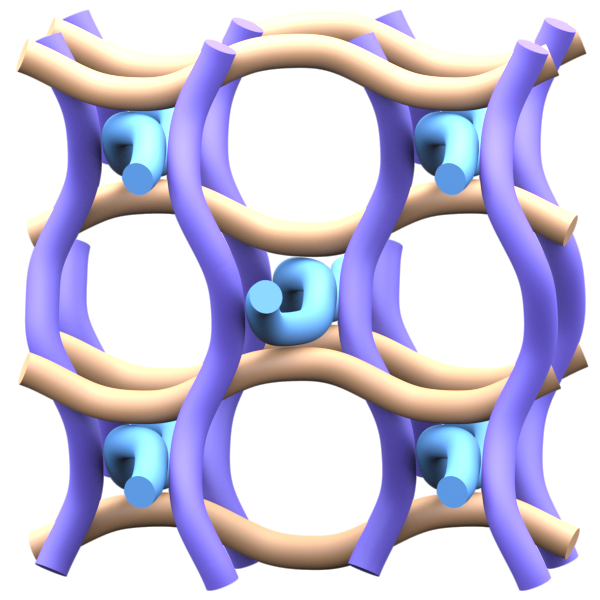}  
    \caption{\textbf{srs} weave $\{\frac{\overline{0.4}}{2} \}^6$}
    \label{SUBFIGURE LABEL 2}
\end{subfigure}
\begin{subfigure}{.32\textwidth}
    \centering
    \includegraphics[width=.95\linewidth]{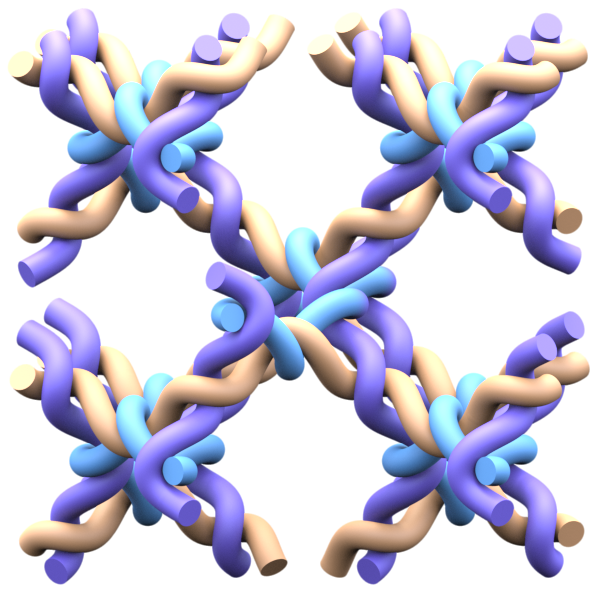}  
    \caption{\textbf{srs} weave $\{\frac{1.6}{2} \}^6$}
    \label{SUBFIGURE LABEL 3}
\end{subfigure}\hfill
\begin{subfigure}{.32\textwidth}
    \centering
    \includegraphics[width=.95\linewidth]{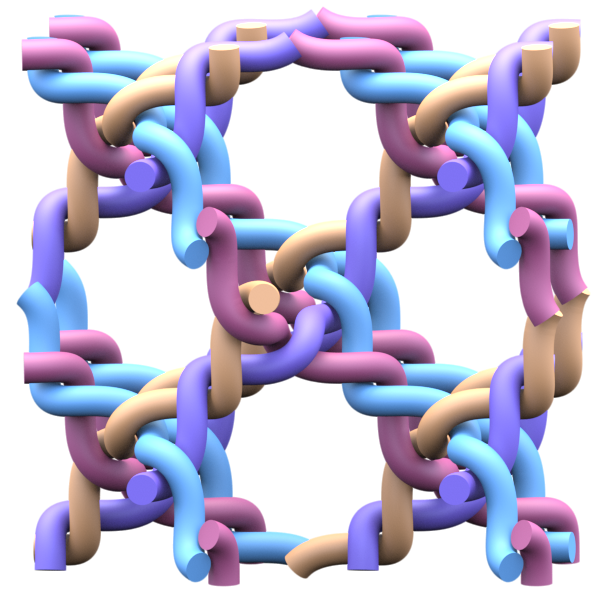}  
    \caption{\textbf{srs} weave $\{\frac{\overline{1.4}}{2} \}^6$}
    \label{SUBFIGURE LABEL 5}
\end{subfigure}
\begin{subfigure}{.32\textwidth}
    \centering
    \includegraphics[width=.95\linewidth]{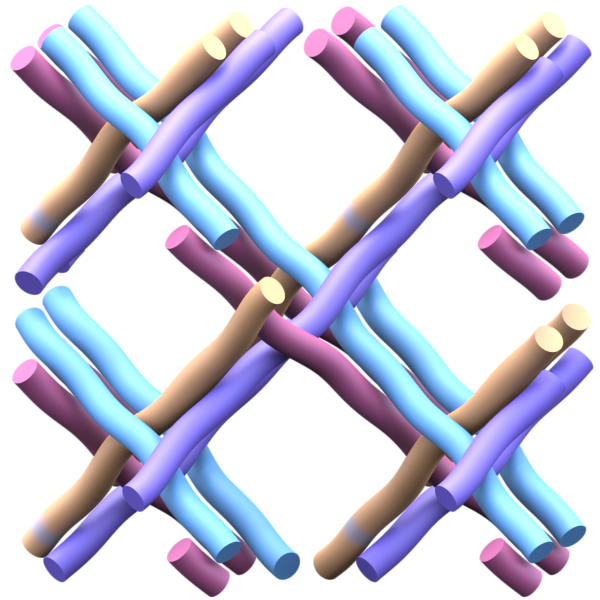}  
    \caption{\textbf{srs} weave $\{\frac{0.6}{2} \}^6$}
    \label{SUBFIGURE LABEL 6}
\end{subfigure}
\begin{subfigure}{.32\textwidth}
    \centering
    \includegraphics[width=.95\linewidth]{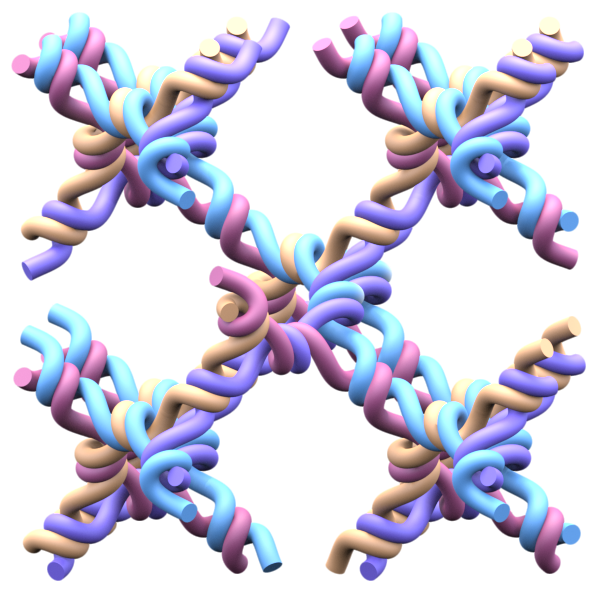}  
    \caption{\textbf{srs} weave $\{\frac{2.6}{2} \}^6$}
    \label{SUBFIGURE LABEL 7}
\end{subfigure}
\caption{Six distinct double-helix tangles hung from the \textbf{srs} network. In all cases, the resulting 3-dimensional structure is a tangling of infinite filaments, where different colours represent set of filaments with invariant axes in the same direction. Structures (a-c) are various tanglings of the $\beta-Mn$ or $\Pi^+$ rod packing \cite{cubic_rod_packings}, where the structure in (b) is the untangled version ambient isotopic to the rod packing itself. Structures (d-f) are tanglings of the $\Sigma^+$ rod packing \cite{cubic_rod_packings}, with structure (e) being the untangled rod packing. All of the structures have $I4_{1}32$ symmetry.}
\label{fig:2helices_weave}
\end{figure}

Figure \ref{fig:2helices_weave} shows a set of simplest examples arising on the left-handed \textbf{srs} net with a weave closure of the double helices. Examples are shown for helical twists of $-2.4\pi$ ($t=-3$), $-1.4\pi$ ($t=-2$), $-0.4\pi$ ($t=-1$), $0.6\pi$ ($t=0$), $1.6\pi$ ($t=1$), and $2.6\pi$ ($t=2$). The resulting structures fall into two categories; those with twist $2j-0.4$ for integer $j$ are tangled versions of the $\beta-Mn$ or $\Pi^+$ rod packings \cite{cubic_rod_packings}, whereas those with twist $2j+0.6$ for integer $j$ are tangled $\Sigma^+$ rod packings \cite{cubic_rod_packings}. Some of these structures have been discussed in the context of hyperbolic tilings previously \cite{Evans:eo5020,myf_ideal_geo}, as well as intermediate filament arrangements in skin \cite{skin_1,PhysRevLett.112.038102}, and exotic mechanical properties \cite{oster2021reentrant, himmelmann_evans_2023_tensegrity}. Some have also been more recently described in the context of reticular chemistry \cite{okeeffe_treacy_2025_3periodic_weavings}. This new construction allows us to extend to a wider variety of tangled structures in a systematic way, pushing into the direction of comprehensive description of symmetric woven materials in three periodic directions.

\begin{figure}[h]
\centering
\begin{subfigure}{.32\textwidth}
    \centering
    \includegraphics[width=.95\linewidth]{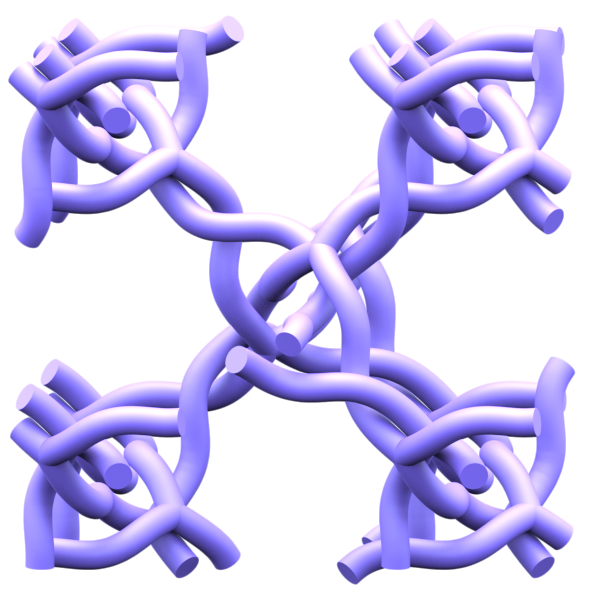} 
    \caption{\textbf{srs} net $\{\frac{\overline{0.9}}{2} \}^6$}
    \label{SUBFIGURE LABEL 1}
\end{subfigure}
\begin{subfigure}{.32\textwidth}
    \centering
    \includegraphics[width=.95\linewidth]{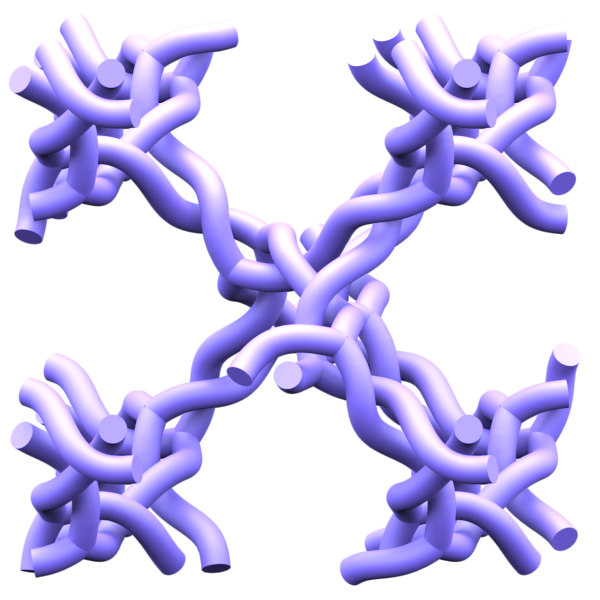}  
    \caption{\textbf{srs} net $\{\frac{1.1}{2} \}^6$}
    \label{SUBFIGURE LABEL 2}
\end{subfigure}
\caption{Two double helical tangles hung from the \textbf{srs} network. Each of the helices placed along the net edges are equivalent, yet the connections of the helices together are done in an alternating fashion, which results in a larger translational unit cell. Both of the structures are tangled single \textbf{srs} networks. Their tangle indices are shown below each structure. Both structures have symmetry $I2_{1}3$.}
\label{fig:2helices_net_weave}
\end{figure}

A curious additional set of examples arise when a combination of net and weave closures are considered. In this case, each double helix in the array must have a net closure end and a weave closure end. This breaks the periodicity of the genus-3 unit cell and requires a larger repeat unit cell for the combinatorics to work, yet only has a very slight reduction in symmetry in the process (space group $I2_{1}3$). We consider these cases here, as they are elegant and interesting. In this case, possible twist values of $t-0.9$ are possible. Figure \ref{fig:2helices_net_weave} shows two such examples with $t=-0.9$ and $t=1.1$. In both cases, the resulting structure is a single, left-handed \textbf{srs} net that is tangled with itself in space, with a different tangling in the two distinct cases. As far as we know, these structure have not been previously described elsewhere. 

\clearpage

\section{Assembling triple helices into tangled structures}

\begin{figure}[h]
\centering
\begin{subfigure}{.24\textwidth}
    \centering
    \includegraphics[width=.95\linewidth]{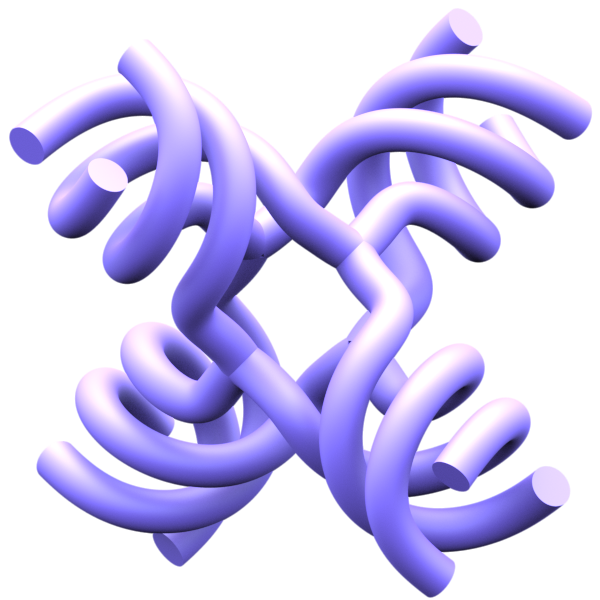}  
    \caption{}
    \label{SUBFIGURE LABEL 3}
\end{subfigure}
\begin{subfigure}{.24\textwidth}
    \centering
    \includegraphics[width=.95\linewidth]{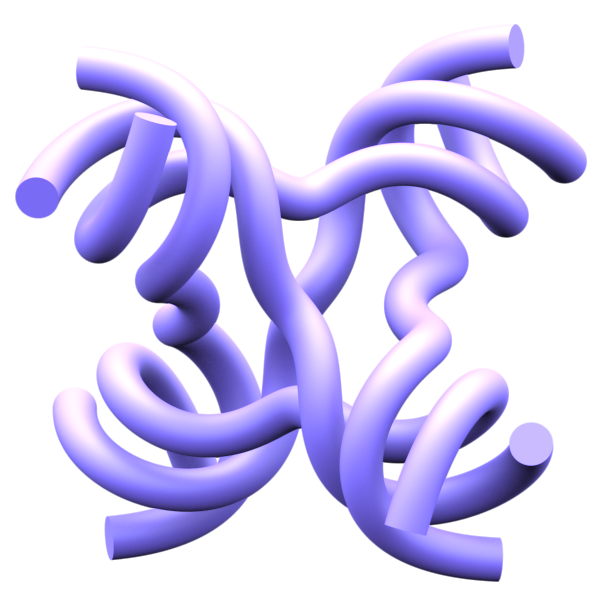}  
    \caption{}
    \label{SUBFIGURE LABEL 4}
\end{subfigure} \hfill
\caption{Two possible maximal symmetry closures of four triple helices arranged around a degree-4 vertex with tetrahedral symmetry. The open ends of the helices can either join together at four degree-3 vertices located at the 3-fold symmetry axes (a), or pairwise through 2-fold symmetry axes (b).}
\label{fig:4-closures}
\end{figure}

The \textbf{dia} net emerges as the natural high-symmetry host for triple helices; the 3-fold rotational symmetry axes along the edges of the \textbf{dia} net allow the placement of triple helices along the edges without a reduction of (rotational) symmetry. The \textbf{dia} net is a genus-3 periodic graph with two distinct vertices and four distinct edges in a periodic unit cell. Along each of these four edges, we can place triple helices with a particular pitch, described by $\frac{t}{3}$, where the 3 strands (denominator) make a twist of $\frac{2t}{3}\pi$ in the helix. The index $\frac{3}{3}$ indicates a full $2\pi$ turn of the double helix. The maximal symmetry closures of the helices results in two specific cases; once again a \textit{net closure} and a \textit{weave closure}. These two closures are shown in Figure \ref{fig:4-closures}, where the net closure results in four degree-3 vertices connecting the helices together, and the weave closure in no vertices but disjoint filaments. Prioritising symmetry dictates that all four triple helices arranged on the \textbf{dia} net should have the same pitch, which gives a \textit{tangle index} of $\{\frac{t}{3}\}^4$.

\begin{figure}[h]
\centering
\begin{subfigure}{.3\textwidth}
    \centering
    \includegraphics[width=.95\linewidth]{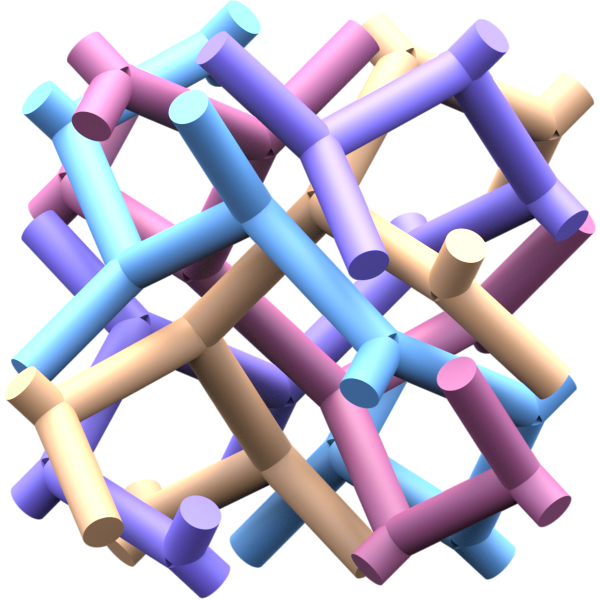}  
    \caption{\textbf{dia} net $\{\frac{0.5}{3} \}^4$}
    \label{SUBFIGURE LABEL 1}
\end{subfigure}
\begin{subfigure}{.3\textwidth}
    \centering
    \includegraphics[width=.95\linewidth]{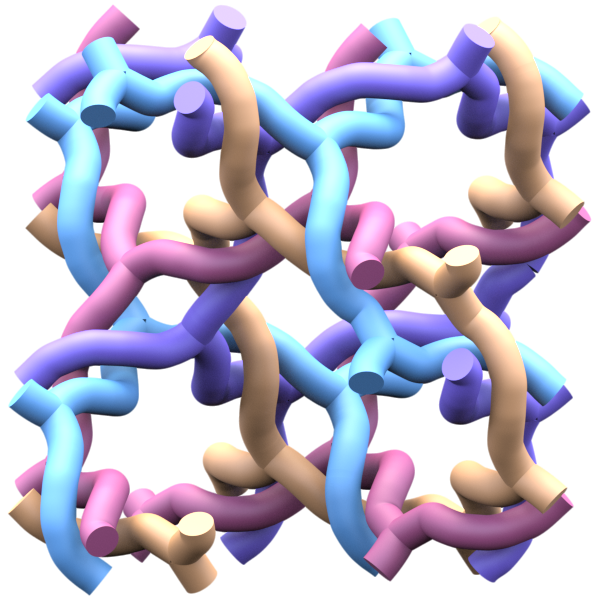}  
    \caption{\textbf{dia} net $\{\frac{1.5}{3} \}^4$}
    \label{SUBFIGURE LABEL 2}
\end{subfigure}
\begin{subfigure}{.3\textwidth}
    \centering
    \includegraphics[width=.95\linewidth]{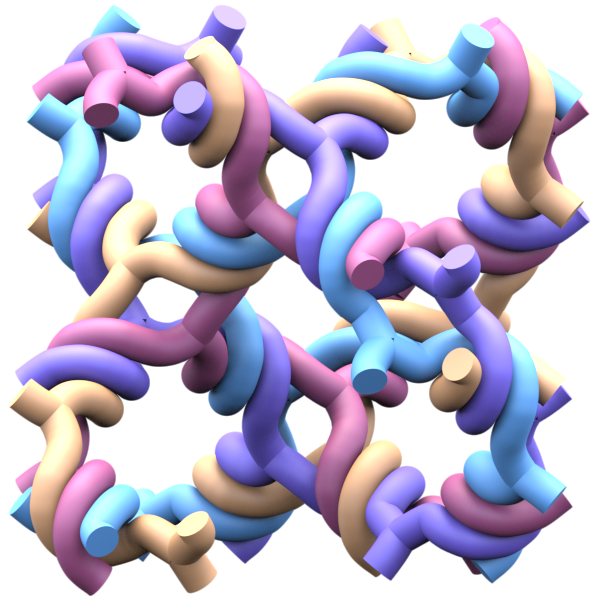}  
    \caption{\textbf{dia} net $\{\frac{2.5}{3} \}^4$}
    \label{SUBFIGURE LABEL 3}
\end{subfigure}
\caption{Three distinct triple-helix tangles hung from the \textbf{dia} network, connected at vertices to give tangled graphs. Structures (a,c) are each the tangling of four \textbf{srs} nets (each coloured differently), where the two structures belong to different ambient isotopy classes. Structure (b) consists of four orientations of interwoven 2-periodic \textbf{hcb} nets in layers, with each orientation coloured differently. All structures have $F4_{1}32$ symmetry.}
\label{fig:3-helices_net}
\end{figure}

With the closures in place, we can now consider the possible values of $t$ in the tangle indices. Neighbouring degree-4 vertices of the \textbf{dia} net are offset from each other; to ensure that helices can join seemlessly at both ends, the twist of the helices must be $t\pm0.5$ where $t$ is an integer. Figure \ref{fig:3-helices_net} shows a set of simplest examples arising on the \textbf{dia} net with a net closure of the triple helices. Examples are shown for helical twists of $0.5$, $1.5$, and $2.5$. Two of the resulting structures contain four distinct but like-handed \textbf{srs} nets in space. The enantiomeric structures can be found using a negative rather than a positive $t$. The simplest of these two entanglements has been described previously; \textbf{dia} net $\{\frac{0.5}{2} \}^4$ is the standard entanglement of four like-handed srs nets \cite{HydeOguey2000,periodic_ent_I}. The other structure (\textbf{dia} net $\{\frac{1.5}{3} \}^4$) is the entanglement of four orientations of 2-periodic hcb nets in layers, also described previously in \cite{periodic_ent_I}.

\begin{figure}[h]
\centering
\begin{subfigure}{.32\textwidth}
    \centering
    \includegraphics[width=.95\linewidth]{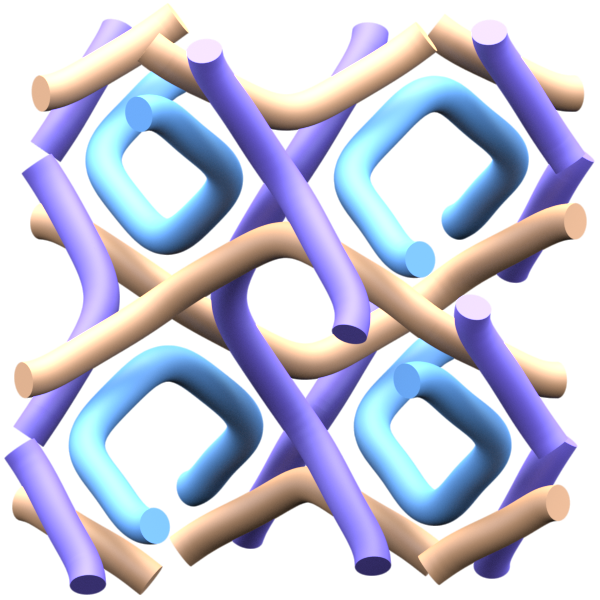}  
    \caption{\textbf{dia} weave $\{\frac{0.5}{3} \}^4$}
    \label{SUBFIGURE LABEL 1}
\end{subfigure}
\begin{subfigure}{.32\textwidth}
    \centering
    \includegraphics[width=.95\linewidth]{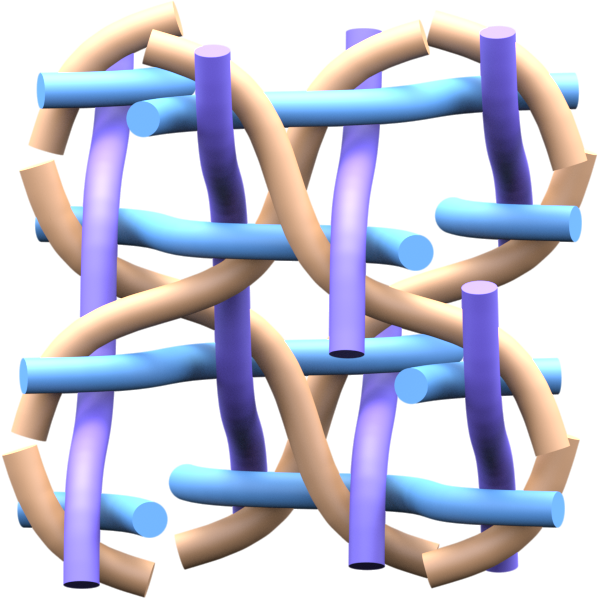}  
    \caption{\textbf{dia} weave $\{\frac{1.5}{3} \}^4$}
    \label{SUBFIGURE LABEL 2}
\end{subfigure}
\begin{subfigure}{.32\textwidth}
    \centering
    \includegraphics[width=.95\linewidth]{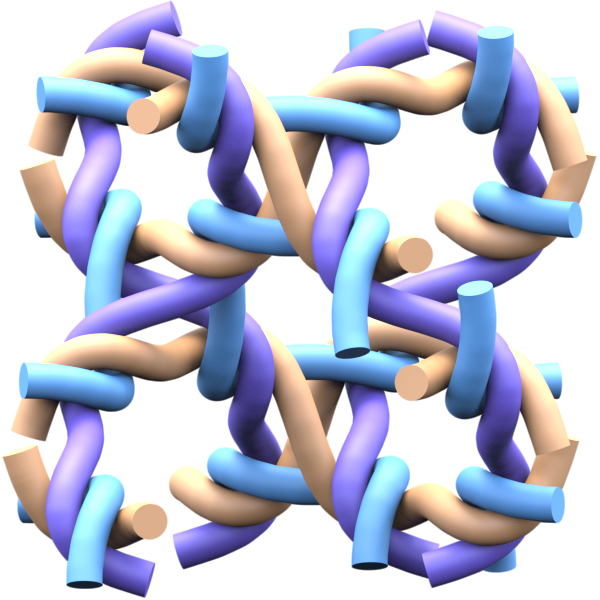}  
    \caption{\textbf{dia} weave $\{\frac{2.5}{3} \}^4$}
    \label{SUBFIGURE LABEL 3}
\end{subfigure}
\caption{Three distinct triple-helix tangles hung from the \textbf{dia} network, connected together to give filament weavings. Structures (a,c) are each the tangling of the $\Pi^*$ or $\beta-W$ rod packing, where the two structures belong to different ambient isotopy classes. Structure (b) consists of six directions of filaments woven together in a symmetric array. All structures have $F4_{1}32$ symmetry.}
\label{fig:3helices_weave}
\end{figure}

Figure \ref{fig:3helices_weave} shows a set of simplest examples arising on the \textbf{dia} net with a weave closure of the triple helices. Examples are shown for helical twists of $0.5$, $1.5$, and $2.5$. Two of the resulting structures are each the tangling of the $\Pi^*$ or $\beta-W$ rod packing \cite{cubic_rod_packings}, where the two structures belong to different ambient isotopy classes. The \textbf{dia} weave $\{\frac{0.5}{3} \}^4$ structure is the untangled version of the rod packing, where the filaments are able to straighten. The other structure consists of six directions of filaments woven together in a symmetric array, previously described in the context of hyperbolic tilings \cite{Evans:eo5020} and reticular chemistry \cite{okeeffe_treacy_2025_3periodic_weavings}.

\begin{figure}[h]
\centering
\begin{subfigure}{.32\textwidth}
    \centering
    \includegraphics[width=.95\linewidth]{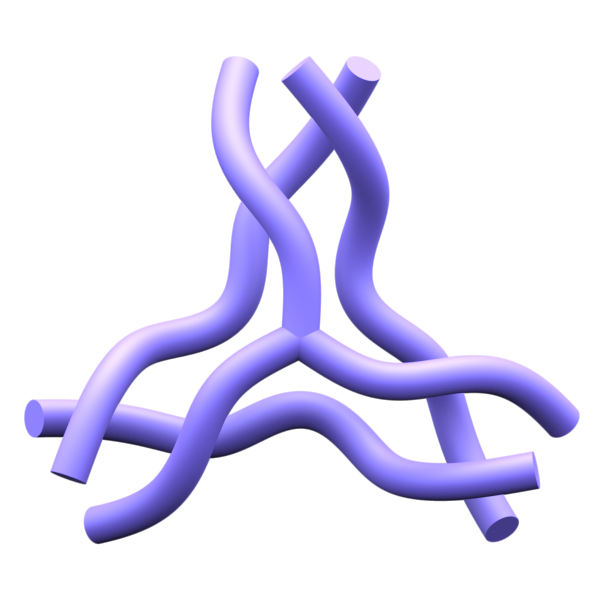}  
    \caption{}
    \label{SUBFIGURE LABEL 1}
\end{subfigure}
\begin{subfigure}{.32\textwidth}
    \centering
    \includegraphics[width=.95\linewidth]{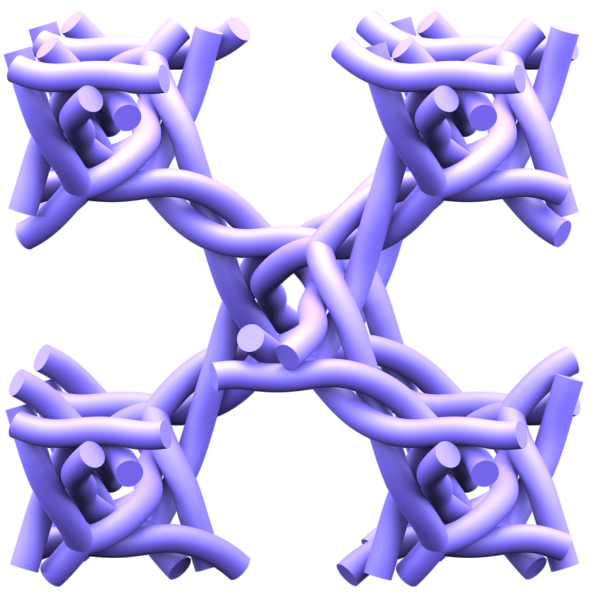}  
    \caption{\textbf{srs} net $\{\frac{\overline{1.1}}{3} \}^6$}
    \label{SUBFIGURE LABEL 1}
\end{subfigure}
\caption{(a) An exotic closure of triple helices at a symmetric 3-fold junction. (b) A triple-helix tangle hung from the \textbf{srs} network. Each of the helices placed along the net edges are equivalent, and have a closure as shown in (a). This results in a larger translational unit cell. The result is a tangled single \textbf{srs} structure with symmetry $I2_{1}3$. }
\label{fig:3helices_fractional}
\end{figure}

Beyond the constructions on the \textbf{dia} net, there is also an interesting case of triple helices that we can construct on the \textbf{srs} scaffold. With only a minor decrease in symmetry, we can create an exotic closure of triple helices to a high symmetry tangled structure on the \textbf{srs} net. The closure is shown in Figure \ref{fig:3helices_fractional}; the breaking of the 2-fold symmetry that maps above the 3-fold vertex to below it results in two parities of closures that are alternated across the vertices of the \textbf{srs} net. This results in a larger periodic repeat unit, much the same as the alternating double helical structures shown in Figure \ref{fig:2helices_net_weave}. The resulting structure is a single, highly tangled \textbf{srs} net, shown in Figure \ref{fig:3helices_fractional}.

\section{Assembling quadruple helices into tangled structures}

\begin{figure}[h]
\centering
\begin{subfigure}{.3\textwidth}
    \centering
    \includegraphics[width=.95\linewidth]{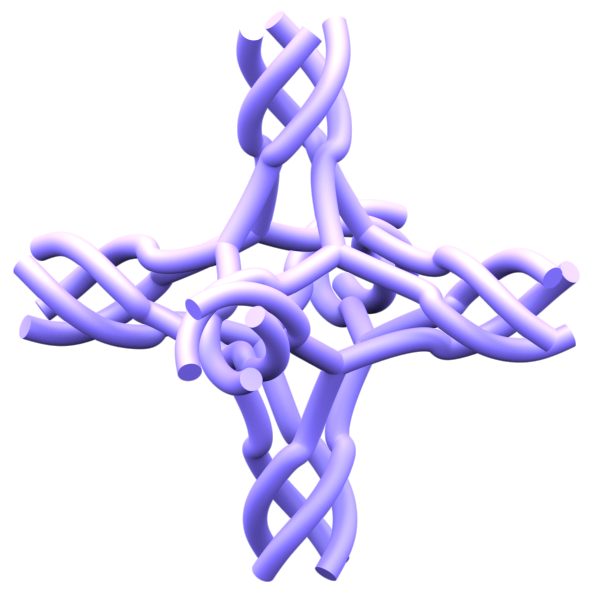}  
    \caption{}
    \label{SUBFIGURE LABEL 3}
\end{subfigure}
\begin{subfigure}{.3\textwidth}
    \centering
    \includegraphics[width=.95\linewidth]{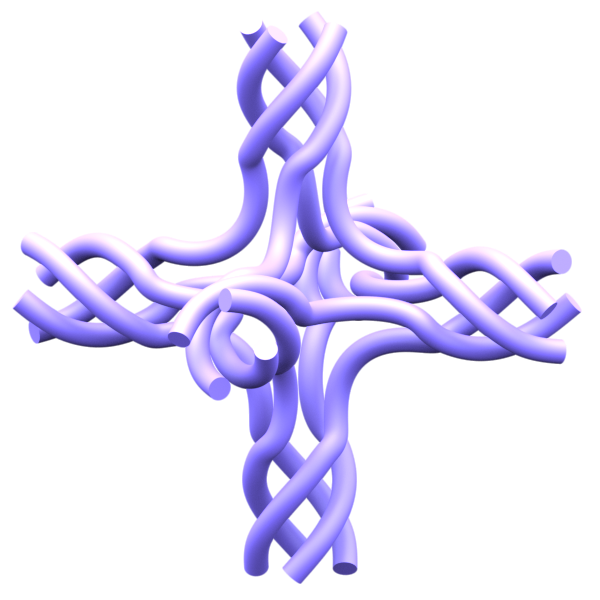}  
    \caption{}
    \label{SUBFIGURE LABEL 4}
\end{subfigure} \hfill
\caption{Two possible maximal symmetry closures of six 4-fold helices arranged around a degree-6 vertex. The open ends of the helices can either join together at eight degree-3 vertices located at the 3-fold symmetry axes (a), or pairwise through 2-fold symmetry axes (b).}
\label{fig:6-closures}
\end{figure}

For 4-fold helices, the geometric requirements shift again; the 4-fold rotational symmetry of the \textbf{pcu} net is compatible with the symmetry of the helix, and also the 2-fold symmetry of the \textbf{srs} net. We consider first the \textbf{pcu} net. The \textbf{pcu} net, with its simple cubic symmetry and explicit 4-fold axes, provides a straightforward framework in which 4-fold helices can be embedded in a highly regular way. The \textbf{pcu} net is a genus-3 periodic graph with one vertex and three distinct edges in a periodic unit cell. Along each of these three edges, we can place 4-fold helices with a particular pitch, described by $\frac{t}{4}$, where the 4 strands (denominator) make a twist of $\frac{t}{2}\pi$ in the helix. There are two possible closures of the open ends of the helices into a closed structure; the \textit{net closure} and the \textit{weave closure} are both shown in Figure \ref{fig:6-closures}. Prioritising symmetry dictates that all three helices arranged on the \textbf{pcu} net should have the same pitch, which gives a \textit{tangle index} of $\{\frac{t}{4}\}^3$.

\begin{figure}[h]
\centering
\begin{subfigure}{.32\textwidth}
    \centering
    \includegraphics[width=.95\linewidth]{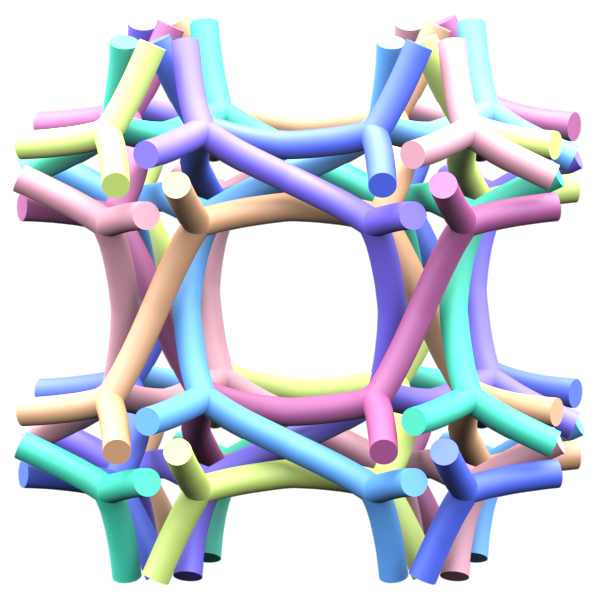}  
    \caption{\textbf{pcu} net $\{\frac{1}{4} \}^3$}
    \label{SUBFIGURE LABEL 1}
\end{subfigure}
\begin{subfigure}{.32\textwidth}
    \centering
    \includegraphics[width=.95\linewidth]{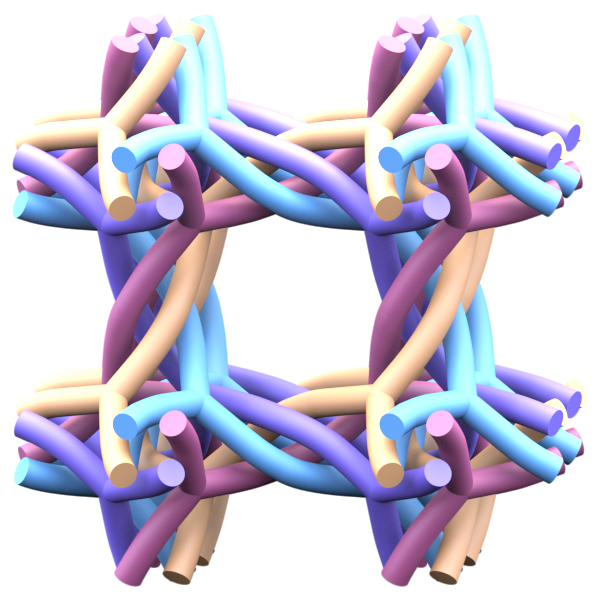}  
    \caption{\textbf{pcu} net $\{\frac{2}{4} \}^3$}
    \label{SUBFIGURE LABEL 2}
\end{subfigure}
\caption{Two distinct 4-fold helix tangles hung from the \textbf{pcu} network. Structure (a) is the tangling of eight \textbf{srs} nets (each coloured differently). Structure (b) consists of four orientations of interwoven 2-periodic \textbf{hcb} nets in layers, with each orientation coloured differently. Both of the structure have space group symmetry $P432$. }
\label{fig:4helices_net_pcu}
\end{figure}

With the closures in place, we can now consider the possible values of $t$ in the tangle indices. In this case, we can now use integer twists, as the closure ends across a helix line up with each other. Figure \ref{fig:4helices_net_pcu} shows two simplest examples arising on the \textbf{pcu} net with helical pitches of $1$ and $2$. The first of these structure is a classical entanglement of 8 like-handed srs nets, as previously described in various contexts \cite{HydeOguey2000,periodic_ent_I}. The second structure is the entanglement of four orientations of 2-periodic hcb nets in layers, also described previously in \cite{periodic_ent_I}.

\begin{figure}[h]
\centering
\begin{subfigure}{.32\textwidth}
    \centering
    \includegraphics[width=.95\linewidth]{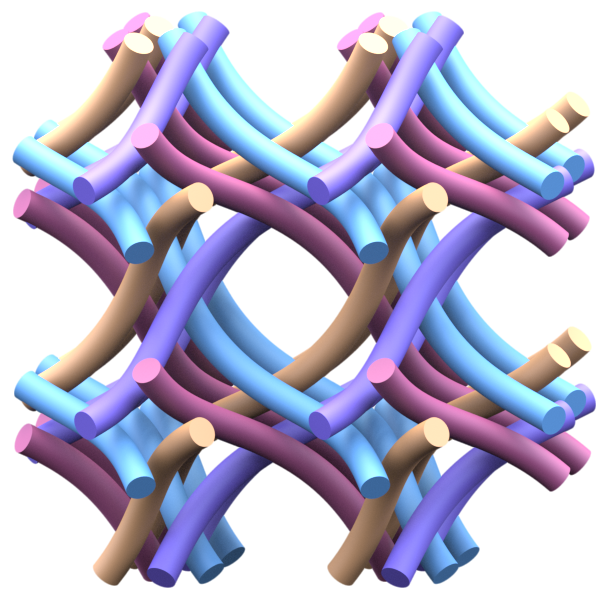}  
    \caption{\textbf{pcu} weave $\{\frac{1}{4} \}^3$}
    \label{SUBFIGURE LABEL 1}
\end{subfigure}
\begin{subfigure}{.32\textwidth}
    \centering
    \includegraphics[width=.95\linewidth]{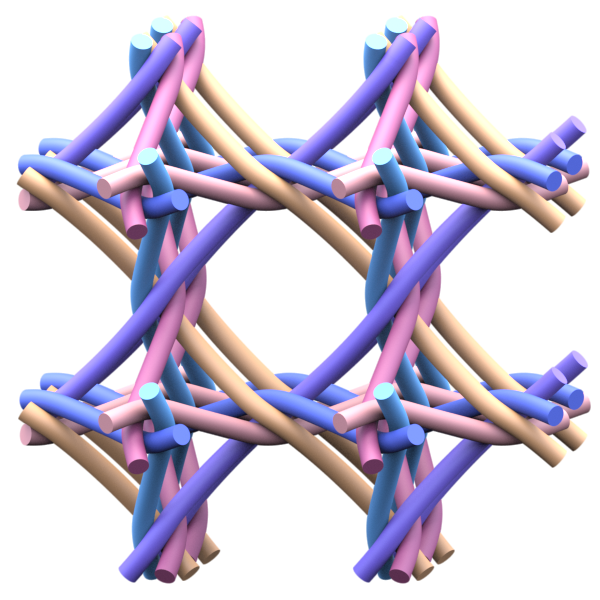}  
    \caption{\textbf{pcu} weave $\{\frac{2}{4} \}^3$}
    \label{SUBFIGURE LABEL 2}
\end{subfigure}
\caption{Two distinct 4-helix tangles hung from the \textbf{pcu} network, connected together to give filament weavings. Structure (a) is the $\Omega^+$ rod packing (with helical filaments). Structure (b) consists of six directions of filaments woven together in a symmetric array. Both of the structure have space group symmetry $P432$.}
\label{fig:6helices_weave_pcu}
\end{figure}

Figure \ref{fig:6helices_weave_pcu} shows two simplest 4-fold helix examples on the \textbf{pcu} net with a weave closure. Examples are shown for helical pitches of $t=1$ and $t=2$. The first structure is the $\Omega^+$ rod packing \cite{cubic_rod_packings}. The other structure consists of six directions of filaments woven together in a symmetric array, as previously describe in the context of hyperbolic tilings \cite{periodic_ent_I}. 

\begin{figure}[h]
\centering
\begin{subfigure}{.3\textwidth}
    \centering
    \includegraphics[width=.95\linewidth]{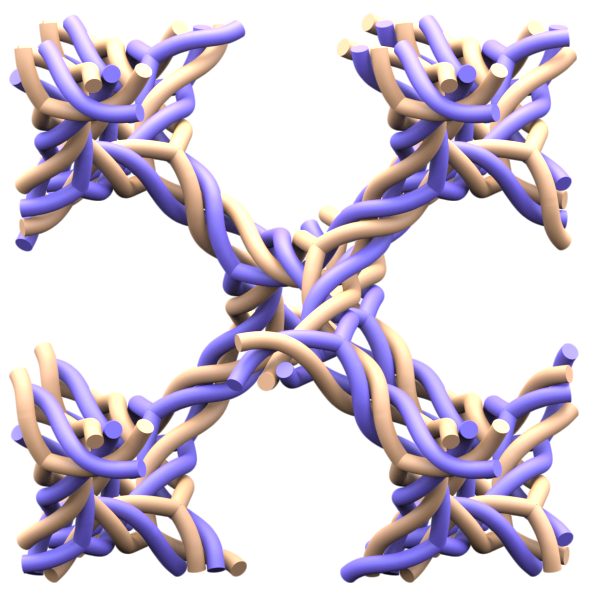}  
    \caption{\textbf{srs} net $\{\frac{2.2}{4} \}^6$}
    \label{SUBFIGURE LABEL 1}
\end{subfigure}
\begin{subfigure}{.3\textwidth}
    \centering
    \includegraphics[width=.95\linewidth]{Figs/SRS_4net_-1p8.png}  
    \caption{\textbf{srs} net $\{\frac{\overline{1.8}}{4} \}^6$}
    \label{SUBFIGURE LABEL 1}
\end{subfigure}
\begin{subfigure}{.3\textwidth}
    \centering
    \includegraphics[width=.95\linewidth]{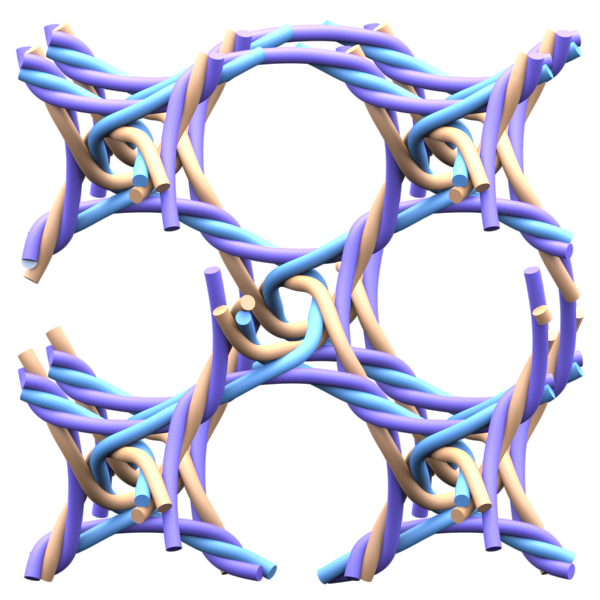}  
    \caption{\textbf{srs} weave $\{\frac{\overline{1.8}}{4} \}^6$}
    \label{SUBFIGURE LABEL 2}
\end{subfigure}
\caption{Three 4-helix tangles hung from the \textbf{srs} network, the first 2 with vertex connections, and the third as filaments.  Structures (a,b) are each the tangling of two \textbf{srs} nets (each coloured differently). In this case they are highly entangled structures. Structure (c) consists of three directions of filaments woven together in a symmetric array. All of these structures have $I4_{1}32$ space group symmetry.}
\label{fig:4helices_srs}
\end{figure}

The \textbf{srs} net is also a good candidate to host 4-fold helical structures. In this case, the closure can be considered as a union of the two 2-fold closures shown in Figure \ref{fig:3-closures}. Two rather elegant structures arise with helical twists of $2.2$ and $-1.8$, as shown in Figure \ref{fig:4helices_srs}. Both of these structures are composed of two like-handed \textbf{srs} nets, which are both self tangled and mutually tangled with each other. An additional structure is also shown in Figure \ref{fig:4helices_srs}, which has quadruple helices with a purely weave closure (a double cover of the weave closure shown in Figure \ref{fig:3-closures}. This woven structure is highly entangled yet elegantly symmetric.

\section{Higher helix numbers and tangle indices}

Beyond the examples shown above, we can also describe structures generated by arrays of helices with higher strand numbers. Generically, these can contain net closures or weave closures, and have a variety of possible tangling indices. These tangling indices can be summarised as follows. A generic structure $\{\frac{t}{n} \}^6$ on the left-handed \textbf{srs} net can be constructed so long as the following rules are followed, for integer $i$ and $j$: $$\{\frac{i-0.4j}{2j} \}^6$$ 

The right-handed srs structures have $+0.4j$ instead. A geometric double cover of the spatial structure results from multiplying the numerator and denominator by 2; for example, the structure $\{\frac{1.2}{4} \}^6$ will be a double cover of the $\{\frac{0.6}{2} \}^6$ structure. Division of the numerator and denominator by 2 also lead to an interesting construction in some cases (including where the structure is not already a double cover of another allowed tangle index). In some cases, division by 2 can result in a structure with slightly decreased symmetry and a larger periodic repeat unit. For example, take the structures \textbf{srs} net $\{\frac{\overline{1.8}}{4} \}^6$ from Figure \ref{fig:4helices_srs} and \textbf{srs} net $\{\frac{\overline{0.9}}{2} \}^6$ in Figure \ref{fig:2helices_net_weave}. The first structure has a form allowed by the generic construction, and the second is the division of both $t$ and $n$ by 2. Geometrically, the second structure is indeed one of the two components from the first structure, reflecting the division by 2 in the tangle index. This demonstrates the elegant relationship between the geometric operations and the tangle indices. 

The indices are also invertible, in the same way that a $\frac{t}{n}$ torus knot can be inverted without changing its knot type. This invertibility corresponds to the helical structure lying on the complementary space in three dimensions. In the case of the srs (or an inflated surface on which the helices lie, the gyroid), one can interchange strands and twist to get a tangle index on the other channel, which is also an srs net of the opposite hand. To do this interchange, one must do the following operation: $$\{\frac{t}{n} \}^6 \rightarrow\{\frac{n/10}{10t} \}^6$$ For example, the structures $\{\frac{\overline{0.4}}{2} \}^6$ (LH \textbf{srs}) and $\{\frac{\overline{0.2}}{4} \}^6$ (RH \textbf{srs}) are indeed the same spatial structure. This invertibility shows powerful algebraic relationship of the indices that reflect geometric operations on the structures.

Structures on the \textbf{dia} net have the form, for integer i and j: $$\{\frac{i-0.5j}{3j} \}^4$$
For these structures, the doubling of the numerator and denominator in the tangle index results in a double cover of the geometric object in space. Some further structures can be achieved through the halving of valid structures combined with an increase in the size of the periodic unit cell. On the diamond, inverting the stands and twists can also be done, reflecting a structure which sits on a complementary \textbf{dia} network. The inversion goes as follows: $$\{\frac{t}{n} \}^4 \rightarrow\{\frac{n/6}{6t} \}^4$$

Analogously, structures on the \textbf{pcu} net can be described with integer twists, so $$\{\frac{i}{4j} \}^3$$
The halving and doubling arguments also hold here. The inversion of the structure is done in the following way: $$\{\frac{t}{n} \}^3 \rightarrow\{\frac{n/4}{4t} \}^3$$

One can see that beyond the simple and elegant structures presented in this paper so far, the technique provides a foundation for a broad and thorough enumeration of many 3-periodic tangled structures, combined with a tangle index descriptor that allows algebraic manipulation in parallel to geometric operations. 

\begin{figure}[h]
\centering
\begin{subfigure}{.35\textwidth}
    \centering
    \includegraphics[width=.95\linewidth]{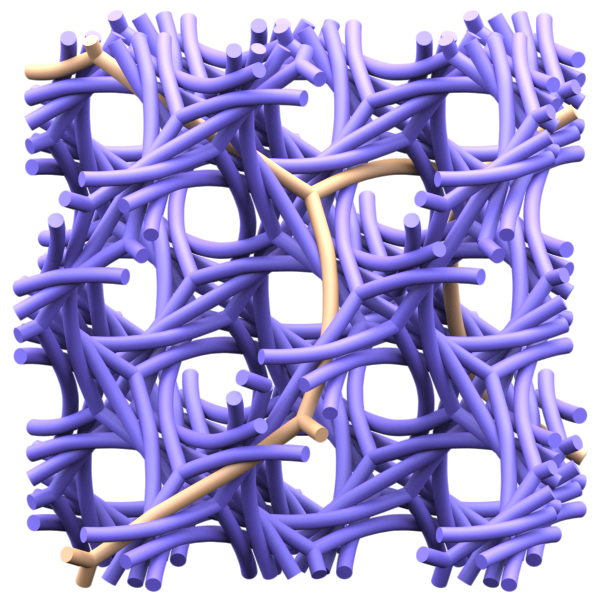}  
    \caption{\textbf{dia} net $\{\frac{2}{6} \}^4$}
    \label{SUBFIGURE LABEL 1}
\end{subfigure}
\begin{subfigure}{.35\textwidth}
    \centering
    \includegraphics[width=.95\linewidth]{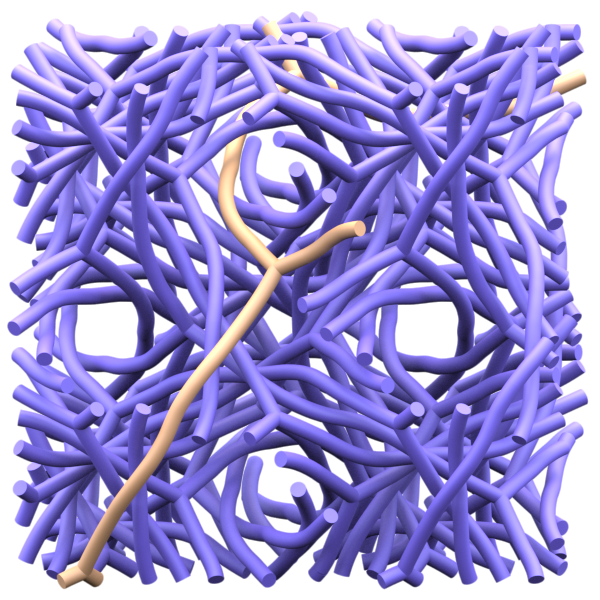}  
    \caption{\textbf{srs} net $\{\frac{1}{10} \}^6$}
    \label{SUBFIGURE LABEL 2}
\end{subfigure}
\caption{Two more complicated structures with higher order helices and higher numbers of network components. (a) A 6-fold helix tangle hung from the \textbf{dia} network, connected at vertices to give tangled graphs, symmetry $F4_{1}32$. It has 64 components in total, where one such component is highlighted in yellow. (b) A 10-fold helix tangle hung from the \textbf{srs} network, connected at vertices to give tangled graphs, symmetry $I4_{1}32$. It has 54 components in total, where one such component is highlighted in yellow.}
\label{fig:higher_helices}
\end{figure}

As an insight into more complicated structure that can be obtained via this framework, we show here two particularly interesting structures in Figure \ref{fig:higher_helices}, with much higher strand numbers of the \textbf{srs} net. These two structures show just how complicated the tangled structures can get, in these cases having 64 components of like-handed \textbf{srs} nets, or 54 components of like-handed \textbf{srs} nets. The second of these structures, \textbf{srs} net $\{\frac{1}{10} \}^6$, has been previously observed in a block-copolymer phase in a computational setting \cite{kirkensgaard2014hierarchical}, and has the elegant property that its inverse structure on the right-handed \textbf{srs} scaffold is indeed also $\{\frac{1}{10} \}^6$, potentially reflecting this balanced optimality of the structure in the simulations. This balanced property is also present in the structures \textbf{dia} net $\{\frac{0.5}{3} \}^4$ (Figure \ref{fig:3-helices_net}) and \textbf{pcu} net $\{\frac{1}{4} \}^3$ (Figure \ref{fig:4helices_net_pcu}).

\section{Conclusions}

This work demonstrates that highly ordered three-periodic tangles can be systematically generated by simple helical motifs. By treating the edges of familiar low-genus periodic nets (notably \textbf{srs}, \textbf{dia}, and \textbf{pcu}) as scaffolds for $n$-fold helices, a rich family of tangled networks and filament weavings emerges under strong symmetry constraints. Despite the apparent complexity of the resulting structures, their organisation follows a small number of geometric principles dictated by rotational symmetry, helix pitch, and connectivity. The introduction of concise tangle indices provides a compact and reproducible description of these embeddings, allowing distinct isotopy classes to be identified and compared within a unified geometric framework.

Beyond serving as a gallery of elegant mathematical constructions, these results highlight tangling as a symmetry-preserving design principle intrinsic to three-dimensional periodic geometry. The structures presented connect naturally to known interpenetrating frameworks, rod packings, block-copolymer morphologies, and biological architectures, reinforcing the idea that entanglement is not incidental but generative in crystalline and soft matter systems. By bridging periodic graph theory, topology, and spatial embedding, this approach complements existing hyperbolic and minimal-surface constructions while offering a flexible route to designing and classifying complex woven materials. More broadly, it clarifies how symmetry, geometry, and topology conspire to produce ordered complexity across scales, from mathematical abstractions to physical matter.

The approach is indeed closely related to the previously published construction idea using hyperbolic tilings printed on the triply-periodic minimal surfaces. Taking the scaffold network as a starting point, the n-fold helices and their closures all lie as non-intersecting lines on a tubified surface around the net. In the case of the nets used in this paper, the tubified surfaces are indeed the Primitive cubic minimal surface (\textbf{pcu}), the Diamond minimal surface (\textbf{dia}) and the Gyroid minimal surface (\textbf{srs}). So the helices of our construction here correspond directly to hyperbolic tree and line packings, replicating many of the previously described structures. What the new construction here provides is a deeper understanding and control of spatial geometry and entanglement, as well as the tangling index that allows clever manipulation of the structures.

We have seen an array of examples from throughout the sciences of these structures arising in experimental or computational settings. In addition to those listed, a related structure has also been observed in the early formation of gyroid photonic crystals in butterfly wingscales \cite{Jessop2025WovenGyroids}. In this case, the variable helix strand number and pitch allows to model a progression of structures which contain helical, tangled filamentous structures. Such examples demonstrate the power of having a variable and broad scheme for the construction of periodic woven geometries with careful constraints on symmetry and underlying net topology.






\funding{This work is funded by the Deutsche Forschungsgemeinschaft (DFG, German Research Foundation) - Project number 468308535. I acknowledge the education licence for Houdini/SideFX, which was used for visualisation.}

\ack{I thank Stephen Hyde for the collaboration on this work, and for the inspiration in seeing the world through geometry, complexity and beauty.}


\printbibliography
\end{document}